# Open comments on the Task Force SIRS report: Scholarly Infrastructures for Research Software (EOSC Executive Board, EOSCArchitecture)

Teresa Gomez-Diaz[‡,§], Tomas Recio[|]

‡ CNRS, Paris, France
§ Laboratoire d'informatique Gaspard-Monge, Est of Paris, France
| University of Nebrija, Madrid, Spain

## Abstract

The goal of this document is to openly contribute with our comments to the EOSCArchitecture report: *Scholarly Infrastructures for Research Software (SIRS)*, and thus, to participate in the European Open Science Cloud (EOSC) architecture design.

## Foreword

The goal of this document is to openly contribute with our comments to the EOSCArchitecture report: *Scholarly Infrastructures for Research Software (SIRS)*. Its draft version was dated October 2020 and it was open for consultation as a Google document until November 10th 2020, as announced at the EOSC Symposium web page (19-22 October 2020, Online)[*1]. It is now an official publication at the EU Publications Office[*2], as announced at the EOSC Architecture group web page[*3]. We refer to this report as "the SIRS report" in the present document.

The first version of our *Open comments on the Task Force SIRS report...* (Gomez-Diaz and Recio 2020b) has been disseminated as a Zenodo preprint dated November 2nd 2020. Now, a few months later, we have decided to present an updated version of our work, including some new reflections motivated by recent information on the SIRS report topics.

Political evolutions in digital policy have been recently announced by Ursula von der Leyen, President of the European Commission (Madiega 2020). They are designed to enhance Europe's strategic autonomy. In this context, and in order to contribute to and to support the Europe's Digital sovereignty, we have avoided the use of Google accounts as much as possible, and in particular, the direct editing of the above mentioned Google document. Therefore, our contribution to this EOSC effort takes the form of the present document.

## Comments

In this section we propose a list of comments to the SIRS report. Each comment is associated to one section or subsection of the report.

Please note that some short extracts of the SIRS report have been included here and are, thus, out of context. We recommend the consultation of the original text, as maybe some mistakes or misinterpretations could have been unintentionally introduced in the present text. These texts correspond now to the official publication and we have updated our comments.

**Section 2.1 Scope and goals - Research software definition** This section 2.1 of the SIRS report introduces the concept of research software used in the report as follows:

> ... *the term "research software" may carry very different meanings in different research communities: in this report, we will use this term simply to designate software that researchers in any discipline may feel the need to have scholarly infrastructure support for, no matter if it is considered a tool, a result or an object of study*.

Please note that this definition does not imply any difference between the concepts of software and research software, as the research software proposed term in the SIRS report could easily include, for instance, any version of the Windows operating system or a commercial scientific software such as Matlab and many other similar products. Moreover, according with this definition proposal, all software developed since 1960 should be considered as research software, as a science history researcher (for example) could feel that it should be preserved as an object for future studies.

It seems to us that to define EOSC infrastructures and services based in this view of research software is a task that requires some essential precisions. Otherwise, using strictly the above research software definition, how could EOSC teams design adapted services? For which target community? In particular, it is necessary to pay careful attention when dealing with software produced by private companies. It seems to us that these questions are not correctly presented or are missing in the report.

On the other hand, The European Open Science Cloud (EOSC) is presented in (European Commission, Commission staff working document 2018) as follows:

> *The EOSC will be a fundamental enabler of Open Science and of the digital transformation of science, offering every European researcher the possibility to access and reuse all publically funded research data in Europe, across disciplines and borders*.

In addition, we have proposed the following definition for Open Science in this recent preprint (Gomez-Diaz and Recio 2020a):

> *Open Science is defined as the political and legal framework where research outputs are shared and disseminated in order to be rendered visible, accessible, reusable*.

According with this Open Science vision, we propose another research software definition as follows:

> *... research software is a well identified set of code that has been written by a (again, well identified) research team. It is software that has been built and used to produce a result published or disseminated in some article or scientific contribution. Each research software encloses a set (of files) that contains the source code and the compiled code. It can also include other elements as the documentation, specifications, use cases, a test suite, examples of input data and corresponding output data, and even preparatory material*.

You can find this definition and all the considerations for its proposition in section 2.1 of (Gomez-Diaz and Recio 2019).

Thus, the research software definition in (Gomez-Diaz and Recio 2019) places correctly research software as a research output, usually produced within publicly funded research, and the role of EOSC efforts correspond to the definition and implementation of infrastructures and services to render research outputs, including research software, *visible, accessible, reusable*.

One of the authors of the present document provides with a good example to further study this research software concept. T. Recio is nowadays studying automatic proving of geometric theorems through dynamic geometry software, and comparing current work with the previously existing one, done with very old software computer programs, such as the computer language Logo[*4] or with more recent ones, such as the Cabri-geometry (commercial) mathematical application[*5]. The outputs of this research are currently being implemented in Geogebra[*6]. Under the SIRS report definition, all these four objects (Logo,

Cabri-geometry, Geogebra, Geogebra developed software by T. Recio and collaborators) should be considered as research software, but, under the definition on (Gomez-Diaz and Recio 2019), only the last one fits well into this concept. Research software is then the result of the ongoing scientific work, and not the historical tool or the commercial software which are used or under study. The output and the producer are therefore correctly identified.

The SIRS report should clarify if historical tools and commercial software should be considered as part of the objects for which EOSC should provide infrastructure and services, and how commercial software should be dealt with.

**Section 2.2 Infrastructures participating in the Task Force (TF)** This section presents summary sheets introducing the nine infrastructures that are represented in the SIRS report: three for the Archives category (HAL, Software Heritage, and Zenodo), three for the publisher category (Dagstuhl, eLife, and IPOL), and three in the aggregators category (OpenAIRE, ScanR, and swMath).

We would like to suggest, for all these nine infrastructures, to add the following information, that could be presented in an homogeneous way:

- Funders, their role: do they provide physical hosting facilities, numerical resources, human resources, other funding... distinguish between public and private funding.
- Governance: with a link to the list of persons involved in the governance and their organization.
- Teams: with a link to the list of the teams involved in the infrastructure and their organization.
- Services: list of services provided.
- Target public: specific research communities, entities...
- Legal framework: links to the texts providing the conditions in which the services are provided, about the personal data processing...
- Software: link to the software that is developed and/or employed to run the platforms and services, including details about their licenses and Free/Open Source Software status.
- Data storage: which are the entities involved in the data storage, under which legislation the storage is realized, and where are the servers that store the whole data (countries, institutions, private companies...).

This Section 2.2 also indicates that:

> *In the context of this report we use the term [...] 'Publishers' are organizations that prepare submitted research texts, possibly with associated source code and data, to produce a publication and manage the dissemination, promotion, and archival process. Software and data can be part of the main publication, or assets given as supplementary materials depending on the policy of the journal. In addition, publishers implement a process for ensuring the quality of the accepted research*

> *material (usually peer Review), which is carried out by a subject-specific community of experts*.

The report could provide further clarification about if these publications include Data papers and Software papers, and if the software mentioned in this paragraph corresponds to the research software defined in the section 2.1 of the SIRS report (see the above comment on this section). For those that are unfamiliar with Data or Software papers, examples of scientific journals publishing biodiversity-related data papers can be found in the Global Biodiversity Information Facility (GBIF) web site[*7]. Examples of scientific journals publishing Software papers are listed in the Software Sustainability Institute (SSI) web site[*8] and some have been analyzed in the section 2.4 entitled *Publication of research software* of (Gomez-Diaz and Recio 2019).

**Section 3.1 Survey on Related Initiatives and Related Works** The SIRS report indicates that:

> ... *it seems that general awareness about the importance of software as a research output has started growing only very recently, around 2010, in particular as a byproduct of the reproducibility crisis (Barnes, 2010; Borgman et al., 2012; Colom et al., 2015; Konrad Hinsen, 2013; Rougier et al., 2017; Stodden et al., 2012)*.

Please note that, at least in France, there have been older initiatives. The PLUME Project (2006-2013), launched by the UREC CNRS unit, has studied research software and its dissemination conditions. It has also published research software descriptions and validated software descriptions (Archimbaud and Romier 2010, Gomez-Diaz 2011, Gomez-Diaz 2015, Gomez-Diaz 2019, Gomez-Diaz and Recio 2019). PLUME has also provided training and support services at national level, see the above mentioned references and the PLUME section *Patrimoine logiciel d'un laboratoire*[*9].

At the time, the term research software was not really existing or maybe its use was not widely extended, so the terms used by the PLUME Project where *logiciel d'un laboratoire*, that is, software produced in a French research lab, and the term *dév. Ens Sup - Recherche* or *dév. ESR*, the short forms of *développements de l'enseignement supérieur et la recherche*, that is, developments realised in the Higher Education andResearch community.

This project was well known of in France, as the members of the PLUME team did a lot of conferences and publications to present the project (Archimbaud and Romier 2010, Gomez-Diaz 2019). In particular, the project was also presented in international conferences like fOSSa 2009[*10].

**Section 3.1.3 Aggregators** The SIRS report indicates that:

> *Another remarkable example is the catalog built by the Plume project in order to collect information about software that is useful for research activities (Plume, 2013): it maintains a collection of over 400 entries manually curated about software*

> *projects that are successfully deployed and in use in at least three different research laboratories*.

Please note that the PLUME Project has published 406 validated software descriptions[*11], 358 research software descriptions in French[*12] and 116 research software descriptions in English[*13] between 2007 and 2013. This catalogue has been produced with publication and peer review procedures (not with automatic aggregation procedures or just manually curated metadata), which have involved 2220 contributors, including writers, reviewers and PLUME team members. The project is frozen and the catalogue is not maintained any more (Archimbaud and Romier 2010, Gomez-Diaz 2019).

**Section 3.2.2 Publishers** The SIRS report mentions that:

> *Over the past few years several publishers have led the effort in the transition towards open access as the predominant model of publication for scholarly outputs. This also paves a path for fair and affordable conditions from the start for the dissemination of software, but support for software outside of specialist journals is still limited*.

As presented in (Directorate-General for Research and Innovation (DG-RTD) of the European Commission and Guédon 2019), the identified four key functions needed by scholarly publishing are:

- **registration** *to establish that work had been undertaken by individuals or groups of researchers at a particular time, and thus their claim to precedence*;
- **certification** *to establish the validity of the findings*;
- **dissemination** *to make scholarly works and their findings accessible and visible*;
- **preservation** *to ensure that the 'records of science' are preserved, and remain accessible, for the long term.*

The SIRS report could clarify what means software support provided by publishers.

Please also note that the section 2.4 of (Gomez-Diaz and Recio 2019) entitled *Publication of research software* does provide another partial panorama of the research software publication world (as defined in this publication). Besides, the section 2.5 *Referencing and citation of research software* provides further studies on the reference and the citation issues for research software.

**Section 3.2.3 Aggregators** The SIRS report mentions here the OpenAIRE Research Graph. Please note that there is also the Software Heritage graph Dataset presented in (Pietri et al. 2019) and funded by the Horizon 2020 research and innovation programme under grant agreement No 825328 (FASTEN). The Software Heritage graph dataset is available in multiple formats, which includes a public instance on Amazon Athena interactive query service for ready-to-use powerful analytical processing.

The OpenAIRE Research Graph was the object of the news at the OpenAIRE end of the year 2020 newsletter[*14], an in its web site[*15] we can find detailed information describing this work. In particular, this research graph is available for download and re-use under a CC-BY license. The OpenAIRE platform provides exploring services as well as an interface to facilitate its enrichment by different research key actors.

The SIRS report could provide further insight over comparisons between these two different graphs and the possible interactions between these two works.

**Section 3.3 Best Practices and Open Problems** This section includes three tables which are surely the result of extended discussions inside the working groups. Please note that for readers that have not been involved in the debates, these tables (and many of the terms and expressions used inside) need further explanation. For example the table in section 3.3.1 Best Practice Principles for Archives includes a last column entitled Priorities indicating levels of Development, Adoption, Research, Harmonization. These terms remain unclear and they could be put in a more precise context.

**Section 3.3.1 Best Practice Principles for Archives** This section indicates that:

> ... *one does not need to reinvent the wheel, the archival community should agree on an overall architecture to integrate existing infrastructures*.

We would like to express our agreement with this idea, which we have found very inspiring. We are not experts in archiving services nor data treatment or management, but it seems to us that archiving software source code may find many common issues with data archiving, and thus, both services could be compared and put into perspective. The SIRS TF team could consult with members of the EOSC-HUB project[*16] like BSC, CSC, CINES (and maybe others) and exchange about common gaps and best practices, in order to not to reinvent the wheel.

In this section we can also find:

> *Last, the ideal architecture interconnecting a variety of infrastructures for research software needs inclusiveness of archives for both open software, as well as non-open software, and the ability to ensure the universal archival and reference of the source code of all software, not just research software*.

As already mentioned in our comment to Section 2.1, the handled definition for research sofware does not provide, in our view, any difference with the concept of software. In particular, it seems difficult to design and build EOSC infrastrutures and services in a way in which there is no difference between sofware produced by privated companies and the research software produced in our University labs. Should EOSC infrastructures ask to private companies to produce references, metadata, and links to related research articles and research data for their produced software?

**Section 3.4.2 Identifiers** This section presents the need of proper identification for software artifacts and presents the SoftWare Heritage persistent identifiers (SWHIDs) as a candidate for solution.

Please note that there are other EOSC teams working on persistent identifiers (PIDs) issues (Directorate-General for Research and Innovation (DG-RTD) of the European Commission and Koers 2020, Directorate-General for Research and Innovation (DG-RTD) of the European Commission et al. 2020) as well as FREYA[*17], a whole EC funded project.

The work proposed in the SIRS report as a solution could be more connected with these others EC funded ongoing efforts, and collaborative work should be carried on in order to propose and adopt consensual solutions.

**Section 3.4.3 Quality and Curation** The table mentions the *Evaluation of source code*.

Please note that the CDUR procedure to evaluate research software has been proposed in (Gomez-Diaz and Recio 2019). CDUR comprises four steps introduced as follows: Citation, to deal with correct RS identification, Dissemination, to measure good dissemination practices, Use, devoted to the evaluation of usability aspects, and Research, to assess the impact of the scientific work.

The SIRS report could consider to include the CDUR procedure in the list of methods to assess research software quality, as CDUR includes the evaluation of the research software source code in the Use step.

**Sections 3.4.4 Metrics, 3.4.5 Guidelines, 3.4.6 Tools and Workflows** The SIRS report could provide further details for proposed guidelines, metrics and tools and workflows.

**Section 4.1.1 Archive** This section mentions:

> *1. Universal archive specifically designed for software source code*
>
> *— proactive archival of all software source code (including all dependencies of research software) [...]*
>
> *2. Scholarly repositories*
>
> *— explicit deposit by identified individuals ...*

In here we find, again, the problem of the definition of research software (see above comments on Sections 2.1 and 3.3.1), and the importance of having well defined objects in order to design sound services and infrastructures to deal with them.

The first item refers to the archiving of every existing software and in the second one we find a more "usual" research software object, as identified individuals do the deposit of their own production in the scholarly repositories. But how the scholarly repositories should deal with the dependencies of the deposited software? Or this should be managed by the identified individuals, mostly researchers?

**Section 4.1.4 Cite/Credit** To cite research software and to give credit to the research software producers is a real issue at stake, we could not agree more. This is why the CDUR proposed research software evaluation procedure dedicates a whole step, the Citation step, to this issue (Gomez-Diaz and Recio 2019), in order to contribute to evolutions of research evaluation and to raise awareness on this matter among researchers and other research key actors.

Moreover, in order to cite research sofware, a reference or citation form should be established by their producers. You can find the description of the PLUME/RELIER software reference cards in the already mentioned presentation of PLUME at fOSSa 2009[*18].

**Section 5.1.1 Interactions** The SIRS report mentions that:

> ... *it is important to ensure a vertical interconnection between an universal software archive and scholarly repositories, for the latter to feed the universal archive (see Figure 5). This requires engineering and funding for the development of proper adaptors*.

The SIRS report could provide further insight on the goals of the Scholarly Infrastructures for Research Software studied in the report concerning their contribution to feed universal archives.

**Section 5.3.1 Advanced Technology Development** The SIRS report mentions *the need of the development of an advanced search engine for software source code*. We think that an Universal Software Archive like the one under study at the SIRS report should equally provide good and sound search interfaces oriented to find and retrieve research software by researchers.

Indeed it is known to be difficult to look for research software (Howison and Bullard 2016). Moreover, research software can be potentially a very interdisciplinary tool, which enhances the difficulty of finding the needed (and already existing) software in order to use it in another discipline context, as it is thus easier to ignore its name or the team that developed it. One solution to facilitate these search engines is in particular provided by the use of good metadata and keywords to tag the research software, as has been proposed in the PLUME platform (Archimbaud and Romier 2010, Gomez-Diaz 2019). Once this Universal Software Archive built, it could be the best place to provide this search and retrieval service if good metadata and keyword tagging are ensured.

## Final Remarks

Finally we conclude this short report with some further questions to be considered by EOSC decision makers, as we think that some of the issues raised by the SIRS report need extended and in-depth reflection.

**Definition** The concept of research software is essential for a sound design of infrastructures and services that will deal with this research output.

**Software Management Plan** It is now widely accepted that a Data Management Plan (DMP) is an important tool when dealing with research data, and DMPs are usually required by funders. Tools for Software Management Plans are also available, see for example (Gomez-Diaz and Romier 2018).

**Services** The SIRS report considers three kind of existing infrastructures: *Archives, Publishers, Aggregators*. The issues studied for software and/or for research software are Archive, Reference, Describe, and Credit, as mentioned in section 2.1.1. On the other hand, if EOSC's goals are to render research software *visible, accessible, reusable*, there is also need for services like, for example, search, testing, and retrieval interfaces. EOSC services should be designed, built and provided in order to answer researchers' needs in the Open Science context (Gomez-Diaz and Recio 2020a).

**User-centric EOSC** As well as the services that will be provided, there is the question of the interactions with foreseen users. Relevant members of the EOSC construction have signed a joint statement[*19] to signal the importance of making EOSC relevant for scientific communities and researchers.

**Architecture** EOSC is already a complex system with several key actors of distinct nature. Interactions and collaborations among all the different components should be designed and developed in order to facilitate the user approach to EOSC.

**Ethical issues** Organisational EOSC Ethics are mentioned in the EOSC Pilot report (EOSC Pilot project 2018):

> ... *insisting on transparency, with strategy and decisions documented and public. It means honesty, including disclosure on financial issues and data usage, so that there is no suspicion of hiding possible conflicts of interest. [...] It should also mean, as discussed below, putting into place systems that support and incentivise the research integrity of individual researchers, and demonstrating a commitment to periodic ethical inspection and oversight by an independent body of experts, acting as an advisory board*.

It also mentions that research integrity has been described as concerned with:

- *Reliability in ensuring the quality of research (by appropriate design, methodology, analysis and the use of resources).*
- *Honesty, in developing, reporting and communicating research in a transparent, fair, full way.*
- *Respect for colleagues, participants, society, ecosystems, heritage and the environment.*
- *Accountability, for the research from idea to publication, for its management and organisation, ...*

**Co-Funding** As mentioned in (Madiega 2020):

> *Reliable digital infrastructure and services are critical in today's society, as the coronavirus crisis has highlighted. A range of initiatives have been proposed or are already under discussion at EU level to accelerate the digitalisation process and enhance Europe's strategic autonomy in the digital field*.

In this context, EOSC decision makers should consider the co-funding of infrastructures already funded by non-EU tech companies in a transparent way.

## Epilog

As far as we understand, there are little differences between the SIRS draft report open for consultation until November 2020 and the official publication that has followed in December 2020. The latter publication states in its p. 6 that:

> *The consultation period ran from October 21 until November 10. All comments received were considered*.

Perhaps some of our comments in *Open comments on the Task Force SIRS report...* (Gomez-Diaz and Recio 2020b) have been considered in order to address the new version, but it seems to us that the differences that we have detected between the two versions of the SIRS report (draft, final publication) rather seem to overlook more or less our propositions.

As we can find in the EOSC Secretariat news[*20] the EOSC Executive Board closes mandate in January 2021 with a final progress report (Directorate-General for Research and Innovation (DG-RTD) of the European Commission et al. 2021). The EOSC Executive Board Work Plan 2019-2020, dated May 2020 (Directorate-General for Research and Innovation (DG-RTD) of the European Commission et al. 2020b) did outline the work to be delivered by the end of 2020. We are thus, at the time to writing this document, at the start of a new EOSC Executive Board two years period, in which, we hope, our open comments contribution could be of some help in the building of this incredible adventure that is the EOSC.

## Acknowledgements

We would like to thank the SIRS TF team for such an inspiring work.

We acknowledge the funding provided by the Laboratoire d'informatique Gaspard-Monge (LIGM) at the University Gustave Eiffel (Est of Paris).

## Author contributions

Both authors have the following roles: Conceptualization, Formal Analysis, Investigation, Methodology, Project Administration, Supervision, Validation, Visualization, Writing – Original Draft Preparation, Writing – Review & Editing.

T. Gomez-Diaz has also participate to Funding Acquisition.

## Conflicts of interest

No competing interests have been detected.

## Endnotes

*1 https://www.eoscsecretariat.eu/eosc-symposium-2020, recording available at https://www.youtube.com/watch?v=U_sxfV0kjEg.
*2 https://op.europa.eu/en/publication-detail/-/publication/145fd0f3-3907-11eb-b27b-01aa75ed71a1/
*3 https://www.eoscsecretariat.eu/working-groups/architecture-working-group
*4 https://en.wikipedia.org/wiki/Logo_(programming_language)

*5　https://swmath.org/software/4928
*6　https://swmath.org/software/4203
*7　https://www.gbif.org/data-papers
*8　https://www.software.ac.uk/resources/guides/which-journals-should-i-publish-my-software
*9　https://projet-plume.org/patrimoine-logiciel-laboratoire. Note that *patrimoine logiciel* translates to the English term *software heritage*.
*10　See fOSSa 2009 archives available at http://fossa2010.inrialpes.fr/, https://www.slideshare.net/fossaconference/presentations, and https://projet-plume.org/ressource/fossa-2009-free-open-source-software-academia-conference-2009-presentations. The PLUME presentation realized by T. Gomez-Diaz is available at https://www.slideshare.net/fossaconference/plume-project-4734924.
*11　https://projet-plume.org/types-de-fiches#logiciel_valide
*12　https://projet-plume.org/fiches_dev_ESR
*13　https://projet-plume.org/en
*14　https://www.openaire.eu/newsletter/archive/278-openaire-end-of-year-2020
*15　https://graph.openaire.eu/
*16　https://www.eosc-hub.eu/partners
*17　https://www.project-freya.eu/en/about/mission
*18　See p. 13 of https://www.slideshare.net/fossaconference/plume-project-4734924.
*19　https://www.openaire.eu/eosc-a-tool-for-enabling-open-science-in-europe. See the statement on the EOSC Secretariat website at https://www.eoscsecretariat.eu/eosc-liaison-platform/post/research-oriented-services-trust-collaboration-sustainability-key.
*20　https://www.eoscsecretariat.eu/news-opinion/eosc-executive-board-closes-mandate-final-progress-report